\documentclass[12pt]{article}
\usepackage{epsfig, amsmath, amssymb}
\usepackage{graphicx,psfrag} 
\newcommand{\be}{\begin{equation}}
\newcommand{\ee}{\end{equation}}
\newcommand{\bea}{\begin{eqnarray}}
\newcommand{\eea}{\end{eqnarray}}
\newcommand{\bA}{\begin{array}}
\newcommand{\eA}{\end{array}}
\newcommand{\bc}{\begin{center}}
\newcommand{\ec}{\end{center}}
\newcommand{\al}{\alpha}

\newcommand{\ra}{\rightarrow}
\newcommand{\del}{\partial}

\newcommand{\ie}{{\it i.e.}}
\newcommand{\eg}{{\it e.g.}}

\newcommand{\Nf}{${\cal N}{=}4$}

\begin{document}


\begin{titlepage}
\vspace{30mm}

\bc

\hfill 
\\         [22mm]

{\huge On Lifshitz scaling and hyperscaling violation \\ [2mm] 
in string theory}
\vspace{16mm}

{\large K.~Narayan} \\
\vspace{3mm}
{\small \it Chennai Mathematical Institute, \\}
{\small \it SIPCOT IT Park, Padur PO, Siruseri 603103, India.\\}

\ec
\medskip
\vspace{40mm}

\begin{abstract}
We explore string/M-theory constructions of holographic theories with
Lifshitz scaling exponent $z$ and hyperscaling violation exponent
$\theta$, finding a range of $z,\theta$-values. Some of these arise as
effective metrics from dimensional reduction of certain kinds of null
deformations of $AdS$ spacetimes appearing in the near horizon
geometries of extremal D3-, M2- and M5-brane theories. The $AdS_5$ 
solution in particular gives rise to $\theta=1$ in $d=2$ (boundary) 
space dimensions. Other solutions arise as the IIA D2- and D4-brane 
solutions with appropriate null deformations, and we discuss the 
phase structure of these systems.
\end{abstract}

\end{titlepage}

{\tiny 
\begin{tableofcontents}
\end{tableofcontents}
}


\section{Introduction}

Gauge/gravity duality \cite{AdSCFT} has enabled fascinating
explorations of strongly coupled quantum field theories, with various
investigations over the last few years exploring non-relativistic and
condensed matter systems \cite{AdSCondmatRev}. These typically have
reduced symmetries compared to anti de Sitter space theories.
An interesting class of theories exhibits Lifshitz scaling symmetry 
of the form\ $t\ra \lambda^z t ,\ \ x_i\ra \lambda x_i ,\ \ 
r\ra \lambda r$, with $z$ the dynamical exponent, and $r$ is the 
radial coordinate in the gravity duals,
\be\label{lifshz}
ds^2=-{dt^2\over r^{2z}} + {dx_i^2+dr^2\over r^2}\ .
\ee
These Lifshitz spacetimes arise in effective gravity theories with 
a negative cosmological constant with abelian gauge fields 
\cite{Kachru:2008yh,Taylor:2008tg}, and in various constructions in 
string theory \cite{Balasubramanian:2010uk,
Donos:2010tu,Gregory:2010gx,Donos:2010ax,Cassani:2011sv,
Halmagyi:2011xh,Singh:2010zs,Narayan:2011az,Chemissany:2011mb} 
(see also earlier work \cite{Li:2009pf,Blaback:2010pp,koroteev,
Azeyanagi:2009pr,Hartnoll:2009ns}).  
A simple subclass \cite{Balasubramanian:2010uk,Donos:2010tu} of such 
constructions involves the dimensional reduction of null deformations 
of $AdS\times X$ spacetimes that arise in familiar brane constructions:
for instance the $AdS_5\times X^5$ null deformation is of the form
\be\label{D3nullLif}
ds^2 = {1\over r^2} [-2dx^+dx^- + dx_i^2 + dr^2]
+ 
g_{++} (dx^+)^2 + d\Omega_S^2\ , 
\ee
with $g_{++}(x^+)$ sourced by one or more fields: the long wavelength 
geometry upon dimensional reduction along the $x^+$-direction 
resembles a $z=2, d=3+1$ Lifshitz spacetime, dual to a $2+1$-dim 
field theory (see \eg\ \cite{Balasubramanian:2011ua} for some recent 
progress on the field theory side). 

Effective gravity theories with abelian gauge fields as well as 
scalar fields \cite{Kachru:2008yh,Taylor:2008tg,Goldstein:2009cv,
Cadoni:2009xm,Charmousis:2010zz,Perlmutter:2010qu,Gouteraux:2011ce,
Bertoldi:2010ca,Iizuka:2011hg,Iizuka:2012iv} are in fact quite rich, 
and have been shown to contain larger classes of solutions exhibiting
interesting scaling properties. In particular, there exist (zero
temperature) metrics with Lifshitz scaling and hyperscaling violation
\be\label{lifhyper}
ds^2 = r^{-2(1-{\theta\over d})} \left( -r^{-2(z-1)} dt^2 + dx_i^2
+ dr^2\right)\ .
\ee
These metrics, rewritten as\ 
$ds^2 = r^{2\theta/d} (-{dt^2\over r^{2z}} + {dx_i^2+dr^2\over r^2})$ 
can be seen to be conformal to Lifshitz spacetimes (\ref{lifshz}). 
Here $d$ is the ``boundary'' spatial dimension (\ie\ the dimension 
of the $x_i$) and  $\theta$ the hyperscaling violation exponent. These 
spacetimes exhibit the scaling
\be\label{lifhyperSymm}
t\ra \lambda^z t ,\quad x_i\ra \lambda x_i ,\quad r\ra \lambda r ,\quad
ds \ra \lambda^{\theta/d} ds\ .
\ee
Various interesting discussions in this context, including condensed 
matter perspectives, appear in \cite{Ogawa:2011bz,Huijse:2011ef}. 
Aspects of holography for these metrics have been discussed in 
\cite{Dong:2012se}. In particular, a basic requirement for obtaining 
physically sensible dual field theories 
is the null energy condition\ $T_{\mu\nu}n^\mu n^\nu \geq 0 ,\ n_\mu n^\mu=0$, 
which gives using the Einstein equations $G_{\mu\nu}=T_{\mu\nu}$,
\be\label{NEC}
(d-\theta) (d(z-1)-\theta) \geq 0\ ,\qquad 
(z-1) (d+z-\theta) \geq 0\ .
\ee

It is interesting to look for configurations in string theory which in
certain limits give rise to effective spacetime descriptions of the
form (\ref{lifhyper}). Indeed \cite{Dong:2012se} already made the
interesting observation that black D$p$-brane supergravity solutions
that arise naturally in string theory give rise to effective Lorentz
invariant ($z=1$) metrics with nontrivial hyperscaling 
violation\footnote{Note that this has parallels with discussions 
in \cite{Perlmutter:2010qu}. We have also been informed that the 
solutions in \cite{Charmousis:2010zz} (similar to (\ref{lifhyper})) 
have string constructions in \cite{Gouteraux:2011ce}, with broken 
scaling related to dimensional reduction.}.
Condensed matter motivations apart, it is useful to explore the space
of possible spacetimes (\ref{lifhyper}) and Lifshitz and hyperscaling
violation exponents $z,\theta$ that arise from string/brane
configurations. With this perspective, we study various classes of
null deformations of $AdS$ spaces in string and M-theory here, and
argue that they give rise to effective metrics (\ref{lifhyper}) with a
range of nontrivial $z,\theta$-exponents.  Some of these (sec.~2)
comprise the dimensional reduction of null normalizable deformations of
the form of $AdS$ shock waves: in this class, the null normalizable
deformation for $AdS_5$ (arising from the extremal limit of D3-brane
stacks) gives rise to a solution with\ $d=2, z=3, \theta=1$:\ this 
is thus in the family $\theta=d-1$, which has been argued to 
correspond to a gravitational dual of a theory containing hidden
Fermi surfaces, as discussed in \cite{Ogawa:2011bz,Huijse:2011ef}.
Others arise from the Type IIA string description of null deformations
of (extremal) M2- and M5-brane solutions in M-theory (sec.~3). These
latter supergravity solutions are best regarded as good descriptions 
in some regime of the full phase structure of these theories along 
the lines of \cite{Itzhaki:1998dd}.

\vspace{3mm}

\noindent \emph{Dimensional reduction:}\ 
In what follows, we will discuss the dimensional reduction of various 
higher dimensional spacetimes to obtain appropriate metrics of the 
form (\ref{lifhyper}), so we state the basic expressions we use. 
Consider a (higher-dimensional) metric\ \ 
$ds^2 = g^D_{\mu\nu} dx^\mu dx^\nu + h(x^\mu) d\sigma_{D_I}^2$ \ 
that we want to dimensionally reduce on the ``internal'' $D_I$-dim 
$\sigma$-space to obtain an effective $D$-dim theory: 
here the warp factor for the internal space depends only on the 
$D$-dim spacetime coordinates $x^\mu$. This has an effective action 
of the schematic form\ 
$S\sim \int d^Dx \sqrt{g^D} h^{D_I/2} (R + \ldots)$:\ to go to the 
effective Einstein frame, we perform a Weyl transformation\ 
$g^D_{E,\mu\nu} = e^{2\Omega} g^D_{\mu\nu}$,\ with\ 
$R_E = e^{-2\Omega} (R + \ldots)$. 
Thus we obtain a $D$-dim spacetime with Einstein metric
\be\label{dimred}
ds^2 = g^D_{\mu\nu} dx^\mu dx^\nu + h(x^\mu) d\sigma_{D_I}^2\ \ \ 
\longrightarrow\ \ \ ds_E^2 = h^{D_I/(D-2)} g_{\mu\nu} dx^\mu dx^\nu\ .
\ee

\section{$AdS_5$ null normalizable deformations and
hyperscaling violation}

The gravity/5-form sector of IIB string theory contains as a solution 
the spacetime 
\be\label{D3nullnorm}
ds^2 = {R^2\over r^2} [-2dx^+dx^- + dx_i^2 + dr^2] + R^2 Q r^2 (dx^+)^2
+ R^2 d\Omega_5^2\ ,\qquad R^4\sim g_{YM}^2N{\al'}^2\ ,
\ee
with no other sources, with $Q$ a parameter of dimension (boundary) 
energy density, and $d\Omega_5^2$ being the metric on $S^5$ (or 
other Einstein space).
Equivalently, the 5-dim part of the metric is a solution to
$R_{MN}=-{4\over R^2} g_{MN}$ arising in the effective 5-dim gravity 
system with negative cosmological constant. This is essentially a
deformation of the familiar $AdS_5\times S^5$ solution arising as the
near-horizon geometry of $N$ D3-branes stacks (in the extremal limit), 
with the boundary metric modification being\ 
$\delta g_{++}\sim {1\over r^2} O(r^4)$.\ From the dual \Nf\ super 
Yang-Mills point of view, this is thus a normalizable deformation 
\cite{Balasubramanian:1998sn} and appears to be a nontrivial state 
of the gauge theory (by comparison, the solutions (\ref{D3nullLif}) 
comprise non-normalizable deformations). These solutions are of the 
form of shock waves in $AdS$ and have been studied elsewhere \eg\
\cite{Janik:2005zt,Grumiller:2008va} (see also \cite{Horowitz:1999gf}). 
This metric (\ref{D3nullnorm}) has also appeared in \cite{Singh:2010zs}, 
as a certain double-scaled ``zero temperature'' limit of a black 
3-brane solution, and some properties of this solution have been 
discussed there.

Here we argue that upon dimensional reduction along a compactified
$x^+$-direction, the resulting metric is conformal to $z=3$ Lifshitz 
spacetimes in bulk $3+1$-dim, the conformal factor giving rise to 
hyperscaling violation. Indeed the 5-dim part of the metric 
(\ref{D3nullnorm}) can be rewritten as (relabelling $x^-\equiv t$)
\be
ds^2 = R^2\left( -{dt^2\over Q r^6} + {dx_i^2 + dr^2\over r^2}
+ Q r^2 \Big(dx^+ - {dt\over Q r^4}\Big)^2\right)\ .
\ee
Then along the lines of \cite{Balasubramanian:2010uk,Donos:2010tu}, we 
regard the $x^+$-direction as compact\footnote{We note that $g_{++}>0$ 
implies that constant $x^-$ surfaces are spacelike while constant 
$x^+$ surfaces are null, somewhat similar to (\ref{D3nullLif}) 
discussed in \cite{Balasubramanian:2010uk,Donos:2010tu}: thus 
$x^-$ is the natural time coordinate.} and dimensionally reduce on 
it, using (\ref{dimred}). This gives the effective (bulk) $3+1$-dim 
Einstein metric
\be\label{liftheta=d-1}
ds^2_E\ =\ (R^2Qr^2)^{1/(4-2)}
R^2 \left( -{dt^2\over Qr^6} + {dx_i^2 + dr^2\over r^2} \right)
=\ {R^3\sqrt{Q}\over r} \Big(-{dt^2\over Qr^4} + dx_i^2 + dr^2\Big)\ ,
\ee
electric gauge field\ $A=-{dt\over Qr^4}$ and scalar\ $e^\phi\sim r$. 
(Closely related solutions have also been discussed in appropriate 
dimensional reductions of certain limits of Schrodinger solutions 
\cite{Balasubramanian:2010uw}.)
We have retained the nontrivial scales $R, Q$ to illustrate their 
higher dimensional origin. This dimensionally reduced metric 
has ``boundary'' spatial dimension $d=2$ and is of the form 
(\ref{lifhyper}) with
\be
2(1-{\theta\over 2}) = 1\ \Rightarrow\ \theta = 1 = d-1\ ,\qquad
2(z-1)=4 \Rightarrow\ z=3\ .
\ee
The family $\theta=d-1$ has been argued to correspond to a 
gravitational dual description of a theory with hidden Fermi 
surfaces \cite{Ogawa:2011bz,Huijse:2011ef} (see also \cite{Dong:2012se}). 
It would thus be interesting to obtain a deeper understanding of 
the present brane configuration.

We note that the higher dimensional metric (\ref{D3nullnorm}) 
exhibits $x^+$-translations and the scaling symmetry
\be
x_i\ra \lambda x_i ,\quad r\ra \lambda r ,\quad x^-\ra \lambda^3 x^- ,\quad
x^+\ra \lambda^{-1} x^+\ ,
\ee
while that of (\ref{liftheta=d-1}) are (\ref{lifhyperSymm}) with 
$d=2,\ z=3,\ \theta=1$. Thus we see 
that the higher dimensional metric exhibits $z=3$ Lifshitz scaling 
in the $t,x_i,r$-subspace, while the hyperscaling violation arises 
from the $x^+$-dimensional reduction. 

From the higher dimensional point of view, the spacetime
(\ref{D3nullnorm}) is asymptotically $AdS_5$: from the lower
dimensional perspective, we have an asymptotically Schrodinger
spacetime arising from the $x^+$-DLCQ of $AdS_5$ in lightcone 
coordinates \cite{Goldberger}. In this context, it
is worth noting that there is in fact a slightly bigger class of
solutions in a gravity-dilaton family with a nonzero $g_{++}$
containing both normalizable and (dilaton $\Phi$ sourced) 
non-normalizable pieces,
\be\label{D3nullr}
ds^2 = {R^2\over r^2} [-2dx^+dx^- + dx_i^2 + dr^2]
+ \Big[{1\over 4} R^2 (\Phi')^2 + QR^2r^2 \Big] (dx^+)^2 + R^2d\Omega_S^2\ ,
\quad\ \Phi = \Phi(x^+)\ .
\ee
In the lower dimensional viewpoint, these interpolate between an 
asymptotic $z=2$ $3+1$-dim Lifshitz spacetime (\ref{D3nullLif}) 
\cite{Balasubramanian:2010uk} for small $r$ and the $z=3, \theta=1$ 
hyperscaling violating metric (\ref{liftheta=d-1}) above for large $r$.
It may be interesting to explore such interpolating solutions 
further: in this context, the metric (\ref{liftheta=d-1}) would 
appear as an effective IR metric with some UV completion.


\subsection{Holographic stress tensor, scalar modes}

The holographic stress tensor \cite{Balasubramanian:1999re,
Myers:1999psa,Emparan:1999pm,de Haro:2000xn,Skenderis:2002wp} for 
these $AdS$ shock-wave-like spacetimes has been discussed in \eg\ 
\cite{Janik:2005zt,Grumiller:2008va}.
To quickly review, consider an asymptotically $AdS$ solution to 
Einstein gravity with negative cosmological constant, with metric 
of the form (we set $R=1$ for convenience here) 
\be
ds^2={dr^2\over r^2} + h_{\mu\nu} dx^\mu dx^\nu
= {dr^2\over r^2} + {1\over r^2} 
\left(g^{(0)}_{\mu\nu}+r^2g^{(2)}_{\mu\nu}+r^4g^{(4)}_{\mu\nu}+\ldots\right) 
dx^\mu dx^\nu \qquad (r\ra 0)\ ,
\ee
in the Fefferman-Graham expansion about the boundary $r=0$. Then 
holographic renormalization methods \cite{de Haro:2000xn,Skenderis:2002wp} 
give rise to relations between the metric coefficients 
$g^{(0)}_{\mu\nu}, g^{(2)}_{\mu\nu}, g^{(4)}_{\mu\nu}, \ldots$, and 
physical observables such as the holographic stress tensor. In 
particular, for a flat boundary metric, we have,
\be
g^{(0)}_{\mu\nu}=\eta_{\mu\nu}\ \ \Rightarrow\qquad  g^{(2)}_{\mu\nu}=0\ ,
\qquad \langle T_{\mu\nu}\rangle = {1\over 4\pi G_5}\ g^{(4)}_{\mu\nu}\ .
\ee
For the $AdS_5$ shock wave spacetime (\ref{D3nullnorm}), this gives\ 
$T_{++} \sim const$. This can be checked directly also 
\cite{Balasubramanian:1999re} by defining the quasilocal stress tensor as\ 
$\tau_{\mu\nu}={2\over\sqrt{h}} {\delta I\over\delta h^{\mu\nu}}$,\ where 
$h_{\mu\nu}$ is the induced boundary metric on the timelike near-boundary 
surface at $r=const$, and\   $I=I_{bulk}+I_{surf}+I_{ct} = 
{1\over 16\pi G_5} \int_{\cal M} d^5x \sqrt{-g} 
(R + 12) - {1\over 8\pi G_5} \int_{\del\cal M} d^4x \sqrt{h} (K + 3) $\ \
is the total action including the surface term and counterterm 
engineered to remove near boundary ($r\ra 0$) divergences (with a 
flat boundary metric), and $K=h^{\mu\nu}K_{\mu\nu}$ is the trace of 
the extrinsic curvature $K_{\mu\nu}$.
Then the quasilocal stress tensor\ 
and 
the gauge theory stress tensor expectation value are 
\be
\tau_{\mu\nu} = {1\over 8\pi G_5} \left(K_{\mu\nu} - K h_{\mu\nu} 
- 3 h_{\mu\nu}\right)\ ,\qquad
{\langle T_{\mu\nu}\rangle} = \lim_{r\ra 0}\ {1\over r^2} \tau_{\mu\nu}\ ,
\ee
where the overall ${1\over r^2}$-factor arises from a regulated 
definition of the (induced) boundary metric.
For (\ref{D3nullnorm}), the only departures from the $AdS_5$ expressions 
are in $\{++\}$-components\footnote{The extrinsic curvature is\
$K_{\mu\nu}=-{1\over 2} (\nabla_\mu n_\nu + \nabla_\nu n_\mu)$,\ where
$n_\mu$ is the outward pointing unit normal to the surface $r=const$. 
With $r=0$ being the boundary here, we have\ $n=-{dr\over r}$ , 
giving\ $K_{\mu\nu}={r\over 2} h_{\mu\nu,r}$.} and we have
\be
K_{\mu\nu}=-{1\over r^2} \eta_{\mu\nu}+Qr^2\delta_{\mu,+}\delta_{\nu,+}\ 
\ \quad \Rightarrow\qquad T_{++}={2Q\over 8\pi G_5}\ ,
\ee
in agreement with the result above.
Thus these shock wave spacetimes correspond to a wave on the boundary 
with nonzero constant energy momentum component $T_{++}$.

Now consider a massless scalar field probe propagating in the 5-dim 
part of the spacetime (\ref{D3nullnorm}): the action\ 
$S=\int d^5x\sqrt{-g} g^{\mu\nu} \del_\mu\phi\del_\nu\phi$\ 
for modes with no $x^+$-dependence ($\del_+\phi=0$) simplifies to
\be
\int d^4x {dx^+\over r^5}\ \left(-Qr^6 (\del_-\phi)^2 +
r^2 (\del_i\phi)^2 + r^2 (\del_r\phi)^2 \right)\ ,
\ee
which is seen to map to that for a scalar in the background 
(\ref{liftheta=d-1}).

\subsection{General $AdS_D$ null normalizable deformations}

Along the lines above, we have the (purely gravitational) $AdS_D$ 
deformation,
\be\label{AdSd+1nullNorm}
ds^2 = {R^2\over r^2} [-2dx^+dx^- + dx_i^2 + dr^2] + R^2Qr^{D-3} (dx^+)^2\ ,
\ee
the $x_i$ being $d$-dim (boundary) spatial coordinates, and $D=d+3$, 
with $Q$ a parameter of dimension energy density in $(D-1)$-dimensions.
This is a solution to\ $R_{MN}=-{D-1\over R^2} g_{MN}$, \ie\ to 
gravity with a negative cosmological constant, and has the 
interpretation of an $AdS_D$ shock wave along the lines of the 
previous sections. In particular, this includes the null normalizable 
deformations of the M2-brane $AdS_4\times X^7$ and M5-brane 
$AdS_7\times X^4$ solutions in M-theory, dimensionally reduced on 
the $X^{11-D}$-space. Recalling that conformal dimensions satisfy 
$\Delta (\Delta-D+1)=m^2R^2$ for $AdS_D$, we see that these are also 
normalizable deformations, the boundary metric being deformed as\ 
$\delta g_{++}\sim {1\over r^2} O(r^{D-1})$.
This metric (\ref{AdSd+1nullNorm}) exhibits the scaling symmetry
\be
x_i\ra \lambda x_i ,\quad r\ra \lambda r ,\quad x^-\ra \lambda^{2+d/2} x^- ,
\quad x^+\ra \lambda^{-d/2} x^+\ .
\ee
Relabelling $x^-\equiv t$, the solution (\ref{AdSd+1nullNorm}) can 
be rewritten as
\be
ds^2 = R^2\left(-{dt^2\over Qr^{D+1}} + {dx_i^2 + dr^2\over r^2} 
+ Qr^{D-3} \Big(dx^+-{dt\over Qr^{D-1}}\Big)^2\right)\ ,
\ee
and dimensionally reduced on the $x^+$-dimension using (\ref{dimred}) 
to obtain
\be\label{lifHypd}
ds_E^2 
=\ {R^2 (R^2Q)^{1/(D-3)}\over r} 
\Big(-{dt^2\over Qr^{D-1}} + dx_i^2 + dr^2\Big)\ .
\ee
The dimensionally reduced metric above has ``boundary'' spatial 
dimension $d=D-3$ and is of the form (\ref{lifhyper}) with
\be
z={d\over 2}+2\ ,\qquad \theta={d\over 2}\ .
\ee
For the special case of $d=2$, this $\theta$ value coincides with 
$\theta=d-1$, as we have seen above.

It is worth discussing the general form of the solutions from the 
lower dimensional point of view (the numerical constants ``$\#$'' 
below can be fixed): the $D$-dim action reduces as
\bea
&& \int d^Dx \sqrt{-g^{(D)}}\ (R^{(D)} - 2\Lambda) \nonumber\\
&& \ \ =  
\int dx^+ d^{D-1}x \sqrt{-g^{(D-1)}}\ (R^{(D-1)} - \#\Lambda e^{-2\phi/(D-3)} 
- \# (\del\phi)^2 - \# e^{2(D-2)\phi/(D-3)} F_{\mu\nu}^2 ) ,\qquad
\eea
where the scalar is\ $g_{DD} = e^{2\phi}$ , the (purely electric) 
gauge field is\ $A=-{dt\over r^{D-1}}$\ and the $(D-1)$-dimensional 
metric undergoes a Weyl transformation as\ 
$g^{(D-1)}_{\mu\nu}=e^{2\phi/(D-3)} g^{(D)}_{\mu\nu}$. It is 
straightforward to check that the solution (\ref{lifHypd}) is 
consistent with the equations of motion, with the scalar of the 
form\ $e^{2\phi}=r^{D-3}$.
These are of the general form of the effective actions studied in
\cite{Goldstein:2009cv,Cadoni:2009xm,Charmousis:2010zz,
Perlmutter:2010qu,Gouteraux:2011ce,Bertoldi:2010ca,Iizuka:2011hg,
Iizuka:2012iv}.

\section{Phases of $AdS$ null deformations in M-theory}

\subsection{M2-branes with null deformations and D2-branes}

Null deformations of $AdS_4\times X^7$ solutions obtained from near 
horizon regions of (extremal) M2-brane stacks in M-theory were 
discussed in \cite{Balasubramanian:2010uk,Donos:2010tu} to obtain 
$z=2$ Lifshitz spacetimes in bulk $2+1$-dimensions\footnote{The 
11-dim supergravity equations are\ 
$R_{MN} = {1\over 12} G_{MB_1B_2B_3} G_N^{B_1B_2B_3} 
- {1\over 144} g_{MN} G_{B_1B_2B_3B_4} G^{B_1B_2B_3B_4}$ ,\ and the 
flux equation\ $d\star G_4+{1\over 2} G_4\wedge G_4=0$, alongwith 
the Bianchi identity for $G_4$: see \eg\ \cite{gauntlettMth} for 
conventions.}. Here we have\footnote{In this entire section, we 
find it convenient to define the radial coordinate $r$ so that 
$r\ra\infty$ is the boundary of the corresponding $AdS$ space.}
\bea\label{AdS4null}
&& ds^2 = {r^4\over R^4} (-2dx^+dx^- + dx_i^2) 
+ {1\over 2} R^2 (\phi')^2 (dx^+)^2 + R^2{dr^2\over r^2} 
+ R^2 d\Omega_7^2\ , \nonumber\\
&&\ \ \ G_{4}={6 r^5\over R^6} dx^+\wedge x^-\wedge dx\wedge dr 
+ C d\phi(x^+)\wedge\Omega_{3}\ ,\qquad R^6\sim Nl_p^6\ ,
\eea
with the scalar $\phi=\phi(x^+)$\ (and $\phi'\equiv {d\phi\over dx^+}$),\ 
$C\sim R^3$ being a normalization constant, and $\Omega_{3}$ is a 
harmonic 3-form on some Sasaki-Einstein 7-manifold $X^7$. 
With a trivial scalar $\phi=const$, this is the $AdS_4\times X^7$ solution.
The conditions $d\Omega_3=0,\ d\star\Omega_3=0,\ d(\star d\phi)=0$,
ensure that the Bianchi identity and the flux equation 
$d\star G_4+{1\over 2} G_4\wedge G_4=0$ are satisfied by the 4-form flux.
In particular, taking $X^7=X^3\times X^4$, and 
$\Omega_3=vol(X^3)$, these are automatically satisfied.

Now let us take the 11-dim circle to be in the $X^4$-space, and study
the IIA description of this M2-brane $AdS_4$-null-deformed system
after dimensional reduction on the 11-th circle. Before we do this, 
let us recall the standard dimensional reduction of M2-branes to 
D2-branes (see \eg\ \cite{Itzhaki:1998dd}),
\be\label{MIIAdimred}
ds^2_{11} = H^{-2/3} dx_{\parallel}^2 + H^{1/3} (dr^2+r^2d\Omega_7^2)\
=\ e^{-2\Phi/3} ds_{10}^2 + e^{4\Phi/3} (dx_{11}+A_\mu dx^\mu)^2\ ,
\ee
where $ds_{10}^2,\ \Phi, A_\mu$ are the IIA string frame metric, dilaton 
and gauge field. With $H\sim {R^6\over r^6}$, we have the M2-branes 
localized in the 8-dim transverse space.
Taking the 11-th dimension to be compact and small, we can take 
$H\sim {N\over r^5}$ to then dimensionally reduce, as discussed in 
\cite{Itzhaki:1998dd}, and obtain the 10-dim D2-brane solution\ ($r$ 
now being the radial coordinate in the seven noncompact transverse 
dimensions).

In the present case, since the null deformation along the $x^+$-direction 
is entirely along the brane worldvolume directions, we expect that 
it simply filters through the dimensional reduction on the 11-th 
circle and appears alongwith $dx_{\parallel}^2$ in the reduced metric. 
To elaborate, the extra metric component $g_{++}$ is unaffected by the 
harmonic function being smeared as\ $H\ra {N\over r^5}$ in the 10-dim 
solution: this extra $g_{++}$ is the only modification induced by the 
null deformation to the standard dimensional reduction of the 
$M2$-branes to $D2$-branes, and gives here a D2-brane solution with 
null deformation.
We then have the 10-dim IIA metric and dilaton
\bea\label{D2null}
ds^2_{st} &=& {r^{5/2}\over R_2^{5/2}} \Big(dx_{\parallel}^2 + 
{R^6(\phi')^2\over r^4} (dx^+)^2\Big) + {R_2^{5/2} dr^2\over r^{5/2}} 
+ {R_2^{5/2}\over r^{1/2}} d\Omega_6^2\ ,\qquad
e^\Phi = g_s {R_2^{5/4}\over r^{5/4}}\ ,\nonumber\\
&& R_2^5 \sim g_{YM}^2 N {\al'}^3\ ,\qquad g_{YM}^2={g_s\over\sqrt{\al'}}\ ,
\qquad \ R^6\sim Nl_P^6\sim g_s R_2^5 \sqrt{\al'}\ ,
\eea
with $r$ now the radial coordinate in the seven noncompact 
transverse dimensions (and we have used the relation 
$l_P=g_s^{1/3} \sqrt{\al'}$ between the 11-dim Planck length, the 
string coupling and the string length). We recall that the scalar 
$\phi$ here arises from the 4-form flux: for $\phi=const$, this is 
the usual D2-brane supergravity solution \cite{Itzhaki:1998dd,
Horowitz:1991cd}, with $F^{(4)}_{+-ir}\sim {r^4\over R_2^5}$. The 
solution (\ref{D2null}) can be checked independently from the IIA 
supergravity equations of motion. Note first that the M-theory 
$G_4$-flux deformation in (\ref{AdS4null}) has no components along 
the 11-th circle and thus reduces in IIA to simply a deformation of 
$F_4=dA_3$. This means that the effective action we need to study is 
simply of the form\ $S_{10}\sim \int d^{10}x \sqrt{-g} 
[e^{-2\Phi} (R + (\nabla\Phi)^2) - |F_4|^2]$, with the modifications 
arising only in the metric and $F_4$. Since the $F_4$ modification 
is lightlike with nonzero $F_{+i_1i_2i_3}$ alone, the equation of 
motion for $F_4$ is automatically satisfied. 
The equations of motion thus differ from those of the usual D2-branes
solution only in\ 
$R_{++}\sim e^{2\Phi} (F_{+ABC} F_+^{ABC}-\# g_{++}F_4^2)$,\
which can be seen to be consistent. 
The resulting 10-dim spacetime is a consistent solution, 
independent of any compactification on the $x^+$-direction.\
The 10-dim Einstein metric here is 
\be\label{D2nullE-10d}
ds_E^2 = e^{-\Phi/2} ds_{st}^2 
= {r^{25/8}\over R_2^{25/8}} \Big( dx_{\parallel}^2 
+ {R^6(\phi')^2\over r^4} (dx^+)^2 \Big)
+ R_2^{15/8} {dr^2\over r^{15/8}} + R_2^{15/8} r^{1/8} d\Omega_6^2\ .
\ee
Keeping the $x^+$-direction noncompact, we dimensionally reduce 
this metric on the $S^6$ using (\ref{dimred})\ (with dimensionless 
conformal factor\ $h={r^{1/8}\over R_2^{1/8}}$, so as to obtain an
effective metric of the right physical dimension): this gives
\be
ds_{E,4d}^2 
= {r^{7/2}\over R_2^{7/2}} (-2dx^+dx^- + dx_i^2)  
+ {R^6(\phi')^2\over R_2^{7/2} r^{1/2}} (dx^+)^2 
+ R_2^{3/2} {dr^2\over r^{3/2}}\ .
\ee
Now for $\phi=const$, we see that this metric is of the form 
(\ref{lifhyper}) with $z=1,\ \theta=-{1\over 3}$ ,\ 
in agreement with \cite{Dong:2012se}.
For $\phi'\neq 0$, let us now consider compactifying the $x^+$-dimension 
to obtain, using (\ref{dimred}), relabelling $x^-\equiv t$, and 
redefining $d\rho\sim r^{-5/2}dr$,
\be\label{M2lifHypd=1}
ds^2_{E,3d} = 
c_1 \rho^{-2} \left( -c_2\rho^{-8/3} dt^2 + dx^2 + d\rho^2 \right)\ ,
\ee
with dimensionful constants $c_1,c_2$.
Now $d=1$ and this is of the form (\ref{lifhyper}) with\
$z={7\over 3} ,\ \theta=0$. This is simply a Lifshitz spacetime with 
no hyperscaling violation. We note that this dimensional reduction is 
not standard Kaluza-Klein reduction, but we expect that the long 
wavelength geometry (\eg\ for zero modes on the $x^+$-circle) is 
of the above form, along the lines of \cite{Balasubramanian:2010uk}.

It is worth mentioning that the 10-dim solution (\ref{D2null}) 
approaches the standard D2-brane solution for large $r$, \ie\ in 
the UV. Far in the UV, the supergravity solution breaks down and 
perturbative 2+1-dim super Yang-Mills theory (with a null 
deformation) is a good description: it would be interesting to 
understand this deformation of the gauge theory better. We recall 
that in the IR, the dual field theory description is expected to be 
a DLCQ of an appropriate lightlike deformation of the M2-brane 
Chern-Simons theory \cite{abjm}.

As a 10-dim solution (\ref{D2nullE-10d}), we see that the size of the 
$x^+$-dimension (Einstein frame, with coordinate size $L_+$) and that 
of the 11-th circle compare as\ 
${R_+\over R_{11}} = {\sqrt{g_{++}} L_+\over e^{2\Phi/3} l_P} 
\sim\ r^{19/48} {R^3\phi' L_+\over R_2^{115/48} l_P}$ .\ Thus the 
$x^+$-circle is large relative to the 11-th circle for $r$ 
sufficiently large: in this intermediate regime, an 
$x^+$-compactification in the 10-dim D2-brane solution appears 
sensible. 

\subsection{M2-branes with null normalizable deformations}

In this case, the $G_4$-flux is the same as for the usual M2-brane 
solution while the metric (\ref{AdSd+1nullNorm}) with $d=3$ can be 
recast as (after re-instating the $X^7$)
\bea\label{AdS4nullnorm}
&& ds^2 = {r^4\over R^4} (-2dx^+dx^- + dx_i^2) + {QR^5\over r^2} (dx^+)^2 
+ R^2 {dr^2\over r^2} + R^2 d\Omega_7^2\ , \nonumber\\
&& G_{4}= {6 r^5\over R^6} dx^+\wedge x^-\wedge dx\wedge dr\ ,
\qquad R^6\sim Nl_P^6\ .
\eea
On the IIA dimensional reduction as described previously, this gives
the 10-dim string frame metric and dilaton for D2-branes with null 
normalizable deformation (with $R_2$ etc defined in (\ref{D2null}))
\be\label{D2nullnorm}
ds^2_{st} = {r^{5/2}\over R_2^{5/2}} \Big(dx_{\parallel}^2 + 
{QR^9\over r^6} (dx^+)^2\Big) 
+ R_2^{5/2} {dr^2\over r^{5/2}} + {R_2^{5/2}\over r^{1/2}} d\Omega_6^2\ ,
\qquad e^\Phi = g_s {R_2^{5/4}\over r^{5/4}}\ .
\ee
This is consistent with the IIA supergravity equations of motion:
the only new piece is\ $R_{++}\sim -e^{2\Phi} g_{++} F_4^2$, which 
can be seen to be consistent.
Dimensionally reducing the 10-dim Einstein metric on the $S^6$ and 
compactifying the $x^+$-dimension, using (\ref{dimred}), we obtain
\bea\label{M2nullnormd=1}
&& ds^2 
= c_1 \rho^{-2/3} 
\left( -c_2 \rho^{-4} dt^2 + dx^2 + d\rho^2 \right)\ ,
\eea
with dimensionful constants\ $c_1={QR^9\over R_2^{16/3}} ,\ 
c_2={R_2^{10}\over QR^9}$ . This effective metric is of the form 
(\ref{lifhyper}) with\ $d=1 ,\ z=3 ,\ \theta={2\over 3}$.

The 10-dim gravity solution breaks down in the far UV, where 
perturbative super Yang-Mills theory is a good description: the null 
normalizable deformation would appear to be a shock-wave-like state 
in the gauge theory.

In the 10-dim Einstein metric, 
the size of the $x^+$-dimension and the 11-th circle compare as\ 
${R_+\over R_{11}} = {\sqrt{g_{++}} L_+\over e^{2\Phi/3} l_P} 
\sim {1\over r^{49/24}} {QR^9\over R_2^{95/24}} {L_+\over l_P}$ .
It thus appears that for $r$ sufficiently small, there exists a 
regime of scales where an $x^+$-compactification in the 10-dim 
solution is sensible.

These solutions thus are of the form of null-deformed D2-brane 
systems, which flow from the $x^+$-dimensional reduction of an UV 
perturbative SYM regime through a IIA supergravity region to a 
11-dim $AdS_4\times X^7$ null deformed phase in the IR.

\subsection{M5-branes with null deformation and D4-branes}

We have the null deformation for the $AdS_7\times X^4$ solution 
($i=1\ldots 4$) obtained from the near horizon region of (extremal) 
M5-brane stacks in M-theory,
\bea\label{AdS7null}
&& ds^2 = {r\over R} [-2dx^+dx^- + dx_i^2] + R^2(\phi')^2 (dx^+)^2
+ R^2 {dr^2\over r^2} + R^2 d\Omega_4^2\ , \nonumber\\
&& G_4 = C vol(X^4) + C'd\phi(x^+)\wedge H_3\ ,\qquad R^3\sim Nl_P^3\ ,
\eea
($C, C'$ being constants) \ie\ $H_3$ is a harmonic form ($dH_3=0 ,
\ d\star H_3=0$), 
and the 11-dim spacetime is of the form $AdS_7\times X_4$, with $X_4$ 
of the form $X_4\equiv X_3\times S^1$. In particular we can take 
$H_3=vol(X_3)$ as the volume form on $X_3$. This thus reduces to 
the effective gravity-scalar system corresponding to an 
$AdS_7$-null-deformation, with the equation\ 
$R_{MN}=-6g_{MN} + {1\over 2} \del_M\phi\del_N\phi , \ \ M,N=\mu,r$ .

The M5-brane solution without any null deformation arises as\ 
$ds_{11}^2=H^{-1/3}dx_{\parallel}^2+H^{2/3}dx_\perp^2$ with 
$H\sim {R^3\over r^3}$ in the near horizon region. 
Using the second equation in (\ref{MIIAdimred}) and dimensionally 
reducing the null-deformed solution (\ref{AdS7null}) on the 11-th 
circle which the M5s wrap, we obtain the 10-dimensional dilaton and 
string frame metric for D4-branes with null deformation
\bea
&& ds_{st}^2={r^{3/2}\over R_4^{3/2}} 
\Big(dx_{\parallel}^2 + {R^3(\phi')^2\over r} (dx^+)^2\Big) 
+ R_4^{3/2} {dr^2\over r^{3/2}} + 
R_4^{3/2} r^{1/2} d\Omega_4^2\ ,\nonumber\\ 
&& \quad e^{\Phi}=g_s{r^{3/4}\over R_4^{3/4}}\ ,\qquad
R_4^3\sim g_{YM}^2N\al'\ ,\qquad g_{YM}^2\sim g_s\sqrt{\al'}\ ,
\eea
This can be seen independently from the IIA supergravity equations too.
We first note that the M-theory $G_4$-deformation above has no components 
along the 11-th circle. Therefore, as before in the case of D2-branes, 
this deformation reduces in IIA to purely a modification of $F_4=dA_3$,
with an effective 10-dim action\ $S_{10}\sim \int d^{10}x \sqrt{-g} 
[e^{-2\Phi} (R + (\nabla\Phi)^2) - |F_4|^2]$, the modifications 
arising only in the metric and $F_4$. Since the $F_4$ modification 
is lightlike with nonzero $F_{+i_1i_2i_3}$ alone, the equation of 
motion for $F_4$ is automatically satisfied. The equations of motion 
thus differ from those of the usual D4-branes solution only in\ 
$R_{++}\sim e^{2\Phi} (F_{+ABC} F_+^{ABC}-\# g_{++}F_4^2)$,\ which can 
be seen to be consistent.
Dimensionally reducing the 10-dim Einstein metric on the $S^4$, 
using (\ref{dimred}), we obtain a 6-dim metric which, for $\phi=const$, 
is of the form (\ref{lifhyper}) with $d=4,\ z=1,\ \theta=-1$, in 
agreement with \cite{Dong:2012se}. 
Now with $\phi'\neq 0$, we compactify the $x^+$-direction obtaining 
the effective 5-dim metric
\bea\label{M5lifHypd=3}
ds_E^2 
=\ c_1 \rho^{-8/3} (-c_2 \rho^{-2} dt^2 + dx_i^2 + d\rho^2)\ ,
\eea
with dimensionful constants $c_1,c_2$. This is of the form 
(\ref{lifhyper}) with $d=3,\ z=2,\ \theta=-1$.\
This is again not standard Kaluza-Klein reduction, but we expect 
the long wavelength geometry to be of the above form, along the 
lines of \cite{Balasubramanian:2010uk}.\ 
We expect a range of scales for the regime of validity of the 
$x^+$-compactification of the 10-dim solution, as before.

\subsection{M5-branes with null normalizable deformations}

The $AdS_7\times X^4$ null normalizable solution (\ref{AdSd+1nullNorm})
can be recast as
\be
ds^2 = {r\over R} [-2dx^+dx^- + dx_i^2] + {QR^8\over r^2} (dx^+)^2
+ R^2 {dr^2\over r^2} + R^2 d\Omega_4^2\ , \qquad
G_4 = C vol(X^4)\ ,
\ee
with $Q$ of dimension energy density in 6-dimensions.
Then after dimensional reduction to IIA, we obtain the 10-dimensional 
dilaton and string frame metric for D4-branes with null normalizable 
deformation
\be
ds_{st}^2={r^{3/2}\over R_4^{3/2}} 
\Big(dx_\parallel^2 + {QR^9\over r^3} (dx^+)^2\Big) 
+ R_4^{3/2} {dr^2\over r^{3/2}} + R_4^{3/2} r^{1/2} d\Omega_4^2\ ,
\qquad e^{\Phi}=g_s {r^{3/4}\over R_4^{3/4}}\ .
\ee
In the IIA supergravity equations, the only new piece is\
$R_{++}\sim -e^{2\Phi} g_{++} F_4^2$ which can be seen to be consistent.
Dimensionally reducing the 10-dim Einstein metric on the $S^4$, 
using (\ref{dimred}), and then compactifying the $x^+$-direction, 
we obtain
\bea\label{M5nullnormd=3}
ds_E^2 = c_1\rho^{-4/3} (-c_2\rho^{-6} dt^2 + dx_i^2 + d\rho^2)\ ,
\eea
with $c_1\sim R_4^{1/3} (QR^9)^{1/3}\ ,\ c_2\sim {R_4^9\over QR^9}$ .
This is of the form (\ref{lifhyper}) with\ 
$d=3,\ z=4,\ \theta={1\over 3}$.

The 10-dim gravity solution breaks down in the IR where 
perturbative super Yang-Mills theory with null deformation is 
expected to be a good description. We expect this to be a 
shock-wave-like state in the gauge theory. In the UV, the description 
is in terms of null deformations of M5-brane $AdS_7\times X^4$ 
solutions, or equivalently null deformations of the dual $(2,0)$ 
superconformal M5-brane theory. It would thus appear that the 
dimensional reduction along the 11-th circle and the $x^+$-direction 
effectively yields a $3+1$-dim nontrivial field theory. It would be 
interesting to understand this better.

\section{Discussion}

We have studied various string/brane configurations and argued that
they give rise to effective metrics of the form (\ref{lifhyper}) with
Lifshitz scaling and hyperscaling violation. The $AdS_5$ null
normalizable deformation (\ref{liftheta=d-1}) corresponds to $d=2,
z=3, \theta=1$, lying in the family $\theta=d-1$, which has been
argued \cite{Ogawa:2011bz,Huijse:2011ef} to be a gravitational dual of
a theory with hidden Fermi surfaces.  Clearly the constructions here
are by no means an exhaustive classification: we expect that there
exist various others too.  We expect that these deformations being
lightlike preserve some supersymmetry since the original brane
solutions themselves are half-BPS: it would be useful to clarify this.

It is interesting to note that the various $z,\theta$-values appearing
in the effective metrics (\ref{liftheta=d-1}), (\ref{lifHypd}),
(\ref{M2lifHypd=1}), (\ref{M2nullnormd=1}), (\ref{M5lifHypd=3}),
(\ref{M5nullnormd=3}), all satisfy the null energy conditions
(\ref{NEC}). This is perhaps not surprising since we are starting with
reasonable matter in string/M-theory. It is worth noting that the null
normalizable deformations have $\theta>0$, while the null
non-normalizable solutions have $\theta\leq 0$:\ it would be
interesting to understand if there is some general correlation
here. It is also worth noting that some of the solutions here, 
\eg\ (\ref{M2lifHypd=1}) (and others with $d=1$), have 
$d-1\leq \theta\leq d$, and thus are expected to have violations of 
the area law for entanglement entropy. We hope to explore these further.

\vspace{5mm}
\noindent {\small {\bf Acknowledgments:} It is a pleasure to thank 
K. Balasubramanian and S. Trivedi for discussions. 
This work is partially supported by a Ramanujan Fellowship, DST, 
Govt. of India.}


\end{document}